\documentclass{emulateapj}
\usepackage{color}
\usepackage{ulem}

\newcommand{\Y}{Y_{lm}(\theta , \phi)}
\newcommand{\diff}{\mathrm{d}}
\newcommand{\Diff}{\mathrm{D}}
\newcommand{\Laplace}{\mathcal{L}}
\newcommand{\M}{\mathcal{M}}

\slugcomment{}

\shorttitle{Growth of Non-spherical Perturbations}
\shortauthors{Takahashi \& Yamada}

\begin{document}

\title{Linear analysis on the growth of non-spherical perturbations in supersonic accretion flows}

\author{Kazuya Takahashi\altaffilmark{1} and Shoichi Yamada\altaffilmark{1,2}}
\affil{$^1$Advanced Research Institute for Science and Engineering, Waseda University, 3-4-1 Okubo, Shinjuku, 169-8555, Japan}
\email{ktakahashi@heap.phys.waseda.ac.jp}
\altaffiltext{2}{Science and Engineering, Waseda University, 3-4-1 Okubo, Shinjuku, 169-8555, Japan}

\begin{abstract}
We analyzed the growth of non-spherical perturbations in supersonic accretion flows. We have in mind the application to the post-bounce phase of core-collapse supernovae (CCSNe). Such non-spherical perturbations have been suggested by a series of papers by Arnett, who has numerically investigated violent convections in the outer layers of pre-collapse stars. Moreover, \citet{Couch} demonstrated in their numerical simulations that such perturbations may lead to a successful supernova even for a progenitor that fails to explode without the fluctuations. This study investigated the linear growth of perturbations during the infall onto a stalled shock wave. The linearized equations are solved as an initial and boundary value problem with the use of Laplace transform. The background is a Bondi accretion flow whose parameters are chosen to mimic the $15$ $\mathrm{M_\odot}$ progenitor model by \citet{Woosley}, which is supposed to be a typical progenitor of CCSNe. We found that the perturbations that are given at a large radius grow as they flow down to the shock radius; the density perturbations can be amplified by a factor of 30, for example. We analytically showed that the growth rate is proportional to $l$, the index of the spherical harmonics. We also found that the perturbations oscillate in time with frequencies that are similar to those of the standing accretion shock instability. This may have an implication for shock revival in CCSNe, which will be investigated in our forthcoming paper in more detail.
\end{abstract}

\keywords{methods: analytical---supernovae: general}

\section{Introduction}
The mechanism of core-collapse supernovae (CCSNe) has been a long-standing problem despite intensive efforts. One of the central issues is how a shock wave proceeds outward and breaks out of an iron core. The shock wave that originates from core bounce does not go through the core directly but stagnates due to heavy accretion and loss of energy via photodissociations of nuclei. Among proposed mechanisms, neutrino heating is commonly thought to be the most promising one, in which neutrinos that diffuse out of a proto-neutron star deposit energy to matter and lead to a revival of the standing shock and, as a result, to an explosion. 

In this scenario, the standing accretion shock instability (SASI) may play an important role to enhance the neutrino heating and contribute to shock revival \citep{Blondin}. SASI is essentially a multi-dimensional instability and some state-of-the-art simulations have confirmed it numerically \citep[see e.g.][and references there in]{Muller, Iwakami, Iwakamib}.
There are also analytical studies on SASI \citep{Yamasaki05, Yamasaki06, Yamasaki07, Foglizzo07, Yamasaki08, Foglizzo09, Sato, Guilet10, Guilet12}. They assumed, however, that the matter flow outside the shock is steady. 

This assumption may not be justified as pointed out by \citet{Bazan, Asida, Meakin06, Meakin07, Arnett, Chat}. They numerically investigated nuclear burnings in outer layers (Si, O, C, and Ne shells) in the pre-collapse stage and found that the structures of progenitors are substantially deviated from spherical symmetry due to violent convections. \citet{Couch} reported recently that such non-spherical fluctuations in the progenitor may yield successful explosions even when no explosion obtains without the fluctuations. Since they computed only a small number of models, comprehensive studies on the effect are awaited.

Instead of conducting multi-dimensional simulations, this paper investigates the growth of non-spherical perturbations in accretion flows onto the standing shock wave, based on linear analysis. In contrast to previous studies \citep[hereafter LG00]{Kovalenko,Lai}, we do not treat the asymptotic behavior $(r\rightarrow 0)$ but deal with the growth of perturbations as an initial and boundary value problem with a use of Laplace transform. This facilitates to see the correspondence between the seed perturbations and the fluctuations at the shock. We employ some assumptions for simplicity: we neglect cooling and heating and use a polytropic equation of state; we consider only the gravity of the proto-neutron star that is approximated by a point mass; the background flow is assumed to be a spherically symmetric supersonic Bondi accretion flow; These assumptions are justified for the current purpose.

In the paper we give perturbations initially at a certain radius, possibly corresponding to Si/O shells, and see how they evolve as they flow inwards. This is in sharp contrast to the ordinary linear analysis. In fact, perturbations do not grow exponentially in time at any fixed point in the current problem. They grow in space. Our analysis is better suited for such problems. We are interested in the amplification factor of perturbations when they reach the shock wave. As shown later in this paper, they are oscillating in time and the typical frequencies are similar to those of SASI. If an analogy with forced oscillations holds true, these amplified perturbations may enhance the SASI activity in turn. The dependence of the amplification factor on $l$, the index of the spherical harmonics $\Y$, is obtained and is found to be different from those claimed by LG00 and \citet{Kovalenko}, who treated the asymptotic regime $(r\rightarrow 0)$.

The paper is organized as follows. In the next section, we give the basic equations with the assumptions mentioned above, and we introduce the Laplace transform method that facilitates the solution of the linearized partial differential equations. We also set the model-parameters in this section. The results are presented in Sec.~\ref{sec.results} and discussions are given in Sec.~\ref{sec.discussion}. Finally, we summarize our findings in Sec.~\ref{sec.summary}.

\section{Method}
As stated in the introduction, we study the evolution of the perturbations that are initially given at a certain radius. Since we have in mind the application to the post-bounce phase of CCSNe, we focus on the amplification factor and its time-dependence at a certain radius downstream, corresponding to the shock position. We study it by linear analysis although the fluctuations may become nonlinear in reality if they grow sufficiently. For such a purpose, Laplace transform is quite useful as shown in section \ref{Laplace method}. We apply it to the linearized equations that govern the evolution of perturbations in supersonic accretion flows and are derived in section \ref{basic eq}. At the end of this section, we introduce models and parameters employed in this paper.

\subsection{basic equations} \label{basic eq}
We consider supersonic accretion flows, which approximate the matter flows outside the standing shock wave in the post-bounce phase of CCSNe. We neglect the self-gravity of the accreting matter and take into account only the gravity of the central accretor, which mimics a proto-neutron star, and of matter inside the shock wave. This is not a bad approximation since the accretor mass is indeed dominant. Since cooling and heating via neutrinos are negligible, we assume that the flows are adiabatic and employ a polytropic equation of state. We note that nuclear burnings, which are also neglected just for simplicity in this study, may actually affect the dynamics. This issue will be addressed elsewhere. 

Under these assumptions, the governing equations are given as follows.
\begin{eqnarray}
\label{eq.cont}
\frac{\partial \rho}{\partial t} +{\bf \nabla }\cdot (\rho {\bf v}) = 0, \\
\frac{\partial }{\partial t}(\rho v_i) + \nabla _j (\rho v_i v_j +\delta _{ij} p) = -\rho \frac{GM}{r^2}\frac{r_i}{r}, \\
\label{eq.eos}
p = K\rho ^\gamma,
\end{eqnarray}
where $\rho $, $p$, ${\bf v}$, $\gamma $, and $K$ are the density, pressure, velocity, ratio of  specific heats and polytropic coefficient, respectively. $G$ and $M$ are the gravitational constant and mass of the central object, respectively.

As repeatedly mentioned, we consider fluctuations to a spherically symmetric, transonic Bondi accretion flow, which is a time-independent solution of Eqs.~(\ref{eq.cont})-(\ref{eq.eos}).
Following LG00, we linearize above equations and express the perturbations as
\begin{eqnarray}
\delta \rho ({\bf r},t) &=&  \delta \rho (r,t) Y_{lm}(\theta , \phi),  \\
\delta {\bf v} ({\bf r},t) &=& \delta v_r (r,t) \Y \hat{{\bf r}} \nonumber \\
 &&+\delta v _\perp (r,t) \left[\hat{{\bf \theta}} \frac{\partial Y_{lm}}{\partial \theta} + \frac{\hat{{\bf \phi}}}{\sin \theta} \frac{\partial Y_{lm}}{\partial \phi} \right] \nonumber \\
 && +\delta v_{rot} (r,t)\left[ -\hat{{\bf \phi}} \frac{\partial Y_{lm}}{\partial \theta} + \frac{\hat{{\bf \theta}}}{\sin \theta} \frac{\partial Y_{lm}}{\partial \phi} \right],
\end{eqnarray}
where $\Y$ is the ordinary spherical harmonics and $\hat{\bf r}, \hat{\bf \theta}$, and $\hat{\bf \phi}$ are unit vectors in spherical coordinates. Then the system of linearized equations is given as
\begin{eqnarray}
\label{rho} \frac{\Diff }{\Diff t}\delta \rho +\frac{\delta \rho}{r^2}\frac{\diff }{\diff r}(r^2v_r)
 +\frac{1}{r^2}\frac{\partial }{\partial r}(\rho r^2 \delta v_r) \;\;\;\; \nonumber \\
-\frac{\rho }{r}l(l+1)\delta v_\perp = 0, \\
\frac{\Diff }{\Diff t}\delta v_r +\frac{\diff v_r}{\diff r}\delta v_r  
 +\frac{\partial }{\partial r}\left(\frac{\delta p}{\rho} \right) = 0, \\
\label{perp}
\frac{\Diff }{\Diff t}(r\delta v_\perp) + \frac{\delta p}{\rho } = 0, \\
\label{rot} 
\frac{\Diff }{\Diff t}(r \delta v_{rot}) = 0, \\
\label{p}
\delta p = \gamma K \rho ^{\gamma -1}\delta \rho,
\end{eqnarray}
where $\Diff /\Diff t := \partial /\partial t +v_r\partial/\partial r$ denotes the Lagrange derivative. 
We note that Eq.~(\ref{rot}) is decoupled from others and can be solved immediately as follows:
\begin{equation}
\label{analy}
\delta v_{rot} (r,t)= \frac{R}{r}\delta v_{rot,R}\left(t -\int _R^r\frac{\diff r'}{v_r(r')}\right) \theta \left( t -\int _R^r\frac{\diff r'}{v_r(r')} \right) ,
\end{equation}
where $R$ is the radius of the outer boundary and the boundary value $\delta v_{rot, R}(t)$ is imposed there. $\theta (t)$ denotes the step function. The solution means that the perturbation $\delta v_{rot}$ is simply advected inwards, increasing its amplitude as $\propto r^{-1}$.

\subsection{Laplace transform} \label{Laplace method}
We introduce here an idea to solve the linearized partial differential equations by Laplace transform. 
Although finite difference methods are more often employed to solve hyperbolic partial differential equations, we prefer Laplace transform. As will be seen below, this method is particularly suitable for our interest: by what factor does the perturbation imposed at a certain point will grow during the advection to a specified point? 

Laplace transform with respective to $t$ is symbolically expressed by an operator $\Laplace $ and defined as follows \citep[e.g.][]{Text}:
\begin{equation}
\label{Laplace}
f^*(s) := \Laplace [f(t)](s) := \int _0 ^\infty f(t) e^{-st} \diff t ,
\end{equation}
where $f(t)$ is a function of $t$, for which Laplace transform exists for some complex number, $s$. 
The advantage in the use of Laplace transform is that the partial differential equations are reduced to ordinary differential equations with respective to $r$ thanks to the relation:
\begin{equation}
\Laplace \left [\frac{\partial f}{\partial t}\right] = s\Laplace [f] -f(0^+),
\end{equation} 
in which the second term on the right hand side is the initial values of $f$ at $t=0$. 
Note that the $r$-dependence is omitted in the above equation for notational simplicity. Laplace transforming the linearized equations (\ref{rho})-(\ref{perp}), we obtain the following equations:
\begin{eqnarray}
\label{Lrho}
&& \frac{\diff }{\diff r}\left( \frac{\delta \rho^* (r, s)}{\rho } \right) \nonumber \\
 && \;\;\; = \frac{1}{1-\M^2} \left\{ \left( \frac{s\M}{c_s} +\M ^2 \frac{\diff}{\diff r} \ln \dot{M}-\frac{\gamma -1}{\rho }\frac{\diff \rho}{\diff r} \right) \frac{\delta \rho^* }{\rho } \right. \nonumber \\
 && \;\;\;\;\;\; \left. +\left[ -\frac{s\M}{c_s} +\M ^2\left( \frac{\diff }{\diff r} \ln \dot{M} -\frac{2}{v_r}\frac{\diff v_r}{\diff r}\right)  \right]
      \frac{\delta v_r^* }{v_r } \nonumber \right. \\
 &&\;\;\;\;\;\; \left. -\M ^2\frac{l(l+1)}{r}\frac{\delta v_\perp^* }{v_r } \right\} , \\
&&\frac{\diff }{\diff r}\left( \frac{\delta v_r^* (r, s)}{v_r } \right) \nonumber \\
 &&\;\;\; = \frac{1}{1-\M^2}\left[ \left( -\frac{s}{v_r} -\frac{\diff}{\diff r}\ln \dot{M} -\frac{\gamma -1}{\rho}\frac{\diff \rho}{\diff r}\right)
       \frac{\delta \rho^* }{\rho }\right. \nonumber \\
 &&\;\;\;\;\;\; \left.  +\left( \frac{s\M}{c_s} -\frac{\diff}{\diff r}\ln \dot{M}+\frac{2\M^2}{v_r}\frac{\diff v_r}{\diff r}\right)
    \frac{\delta v_r^* }{v_r } \right. \nonumber \\
 &&\;\;\;\;\;\; \left. +\frac{l(l+1)}{r} \frac{\delta v_\perp^* }{v_r } \right] , \\
\label{Lperp}
&& \frac{\diff }{\diff r}\left( \frac{\delta v_\perp^* (r, s)}{v_r } \right) \nonumber \\
 &&\;\;\; = -\frac{1}{r\M^2}\frac{\delta \rho^* }{\rho } \nonumber \\
 &&\;\;\;\;\;\;  -\left( \frac{s}{v_r} +\frac{1}{r} +\frac{1}{v_r}\frac{\diff v_r}{\diff r}\right) \frac{\delta v_\perp^* }{v_r },
\end{eqnarray}
where variables with a suffix $*$ are the quantities that are Laplace transformed with respect to $t$; $\dot{M} (:= 4\pi r^2\rho v_r)$ is the mass accretion rate; $c_s (:= \sqrt{\gamma p /\rho})$ is the sound speed and $\M (:= v_r/c_s)$ is the Mach number of the unperturbed flow. In deriving these equations, we assume that the initial perturbations are zero except at the outer boundary because we suppose that the fluctuations are initially confined in the convective zone in the outer envelope of the progenitor and will fall onto the stalled shock wave later.

We emphasize here that Eqs.~(\ref{Lrho})-(\ref{Lperp}) form a system of ordinary differential equations with respect to $r$ with $s$ being a parameter. 
The integration of the equations is then much facilitated by the use of e.g. the Runge-Kutta method. 
As mentioned above, perturbations are generated at the outer boundary and given as the boundary condition there in our formulation. For example, if the perturbation is given as $f(t,R) = \sin (\omega t)$, then its Laplace transform, $\Laplace [\sin(\omega t)] = \omega /(s^2 +\omega ^2)$, is used as the boundary value for each $s$.
  
Integrating the Laplace transformed equations from the outer boundary to the inner boundary, which corresponds to the shock radius in the CCSNe context, we obtain the Laplace transformed quantities at the inner boundary for a given value of $s$. Collecting these quantities for a set of $s$, we can recover the corresponding time evolutions of these variables at the inner boundary via the inverse transform formula:
\begin{eqnarray}
\label{inverse}
f(t) &=& \Laplace ^{-1}[f^*(s)] ,\\
\label{inverse_x}
 &=& \lim _{y \to \infty}\frac{1}{2\pi i}\int _{x -iy} ^{x+iy} f^*(s) e^{st} \diff s, \\
\label{inverse2}
 &=& \frac{e^{tx}}{2\pi}\int _{-\infty}^\infty f^*(x,y) e^{iyt} \diff y,
\end{eqnarray}
where $x$ and $y$ are the real and imaginary parts of $s$, respectively.\footnote[3]{As a matter of fact, we can obtain the time evolution at any point in the same way.} In these expressions, $x$ is a fixed number and the integral path is a line parallel to the imaginary axis. In fact, $x$ is arbitrary as long as the Laplace transform of $f(t)$ is defined for $s$.
Note that the integral on the right hand side is nothing but a Fourier transform and can be performed efficiently by Fast Fourier Transform. It is also mentioned that if $f(t)$ is a real function, the real and imaginary parts of $f^*(s)$ are even and odd functions, respectively. This is indeed the case for our current problem, since we are dealing with the perturbations of real quantities. We can then reduce the integral domain by half. In the numerical evaluations, the integral domain is enlarged until we see a convergence. Other technical details in numerical evaluations are given in Appendix \ref{appA}.

\subsection{models and parameters}\label{models}
As mentioned repeatedly, we have in mind the application of our models to the post-bounce phase of CCSNe. We hence employ the transonic, spherical Bondi accretion flow as an unperturbed state in order to mimic the infall of the outer envelope onto the stalled shock wave. Only the supersonic portions of the transonic flows are adopted in our models. 
We set the inner boundary at $200-400$ $\mathrm{km}$ from the center, roughly corresponding to the radius of the stagnant shock wave in the post-bounce core of CCSNe. The outer boundary approximately coincides with the position of Si/O layer.

\begin{figure}[ht]
\includegraphics[bb = 0 0 640 480, width = 85mm]{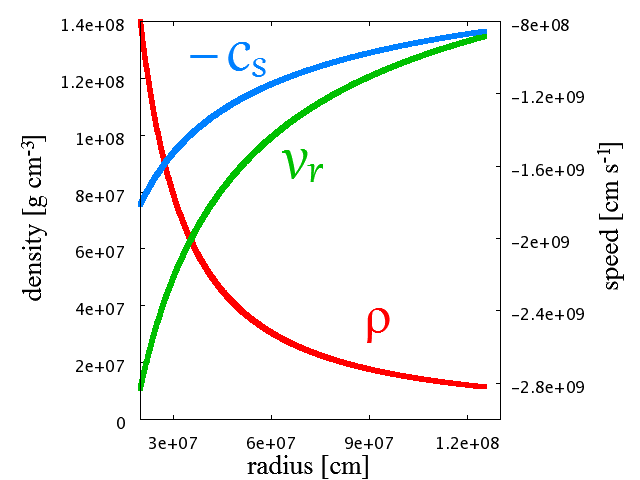}
\caption{The background flow (the transonic Bondi accretion flow) for the canonical parameter set. The blue, green, and red lines represent the sound speed (multiplied by $-1$), flow velocity and density, respectively.}
\label{fig.background}
\end{figure}

The canonical values of model parameters to specify a background flow are as follows: the density at the sonic point $\rho _c = 1\times 10^7$ $\mathrm{g\ cm^{-3}}$, mass accretion rate $\dot{M} = 1$  $\mathrm{M_\odot \ s^{-1}}$, ratio of specific heats $\gamma = 1.6$, mass of the accretor $M_{cen} = 1.4$ $\mathrm{M_\odot}$, and radius of the inner boundary $r_{sh} = 3 \times 10^7$ $\mathrm{cm}$. In this model, the radius of the sonic point is given as $r_c = 1.39 \times 10^8$ $\mathrm{cm}$, and the sound speed at that point is $c_{sc} = 8.20 \times 10^8$ $\mathrm{cm\ s^{-1}}$, and the Bernoulli constant is $E = 1.12 \times 10^{17}$ $\mathrm{g\ cm^2\ s^{-2}}$. The constant in the polytropic equation of state is given as $K = 2.65\times 10^{13}$ in $\mathrm{CGS}$ units. These parameters are tuned to approximate the collapse of $15$ $\mathrm{M_\odot}$ progenitor model by \citet{Woosley} \citep[see also][]{Yamamoto}, which is thought to be a typical progenitor of CCSNe and commonly used in the literature. The profiles of this model are displayed in Fig.~\ref{fig.background}. Note that the mass of accreting matter is $0.1\ \mathrm{M_\odot}$, which is much smaller than the mass of central accretor, $1.4\ \mathrm{M_\odot}$. This justifies the neglect of self-gravity in our model. In fact, the background flow changes at most 10 per cent if we take fully into account the self-gravity of accreting matter.

In reality, the infall velocities around Si/O layer, the region that produces perturbations, will be subsonic. We do not include the subsonic part of the Bondi accretion flow in this study, however, to avoid numerical complexities in treating the sonic point. Unless the perturbations are suppressed substantially in this subsonic portion, which is rather unlikely, the conclusion of this paper is not changed.

As mentioned already, we impose perturbations at the outer boundary as a time-dependent boundary condition. Since the background flow is supersonic, no boundary condition is needed at the inner boundary. As for the functional form of the outer boundary condition,
we consider a step-function, $\theta(t)$, and a sinusoidal function, $\sin (\omega t)$.

\section{Results}\label{sec.results}
In this section, we show the time evolutions of the perturbations as well as their systematics. As mentioned previously, the results are obtained by integrating the Laplace transformed equations (\ref{Lrho})-(\ref{Lperp}) numerically from the outer boundary located near the sonic point to the inner boundary for a set of $s$ and then inverse-transforming the quantities so obtained into the counterparts in the real-time domain by Eq.~(\ref{inverse}). In the following we first present the time evolutions for the step-function type outer boundary condition, for which the causality is demonstrated most clearly. Then we give a more realistic case, in which we impose a sinusoidal time-variations in all quantities at the outer boundary.

\subsection{step-function type perturbations} \label{result.step}
We present firstly the time evolutions at $r_{sh} = 300$ $\mathrm{km}$ in the canonical model. The density of perturbations and radial-velocity perturbations are set at the outer boundary $r = R = 1.25\times 10^{8}$ $\mathrm{cm}$ as a step function in time. Since the background flow is spherically symmetric and the perturbations are decomposed with the spherical harmonics, different $l$-modes are considered separately.

\begin{figure}[htb]
\includegraphics[bb = 0 0 640 480, width = 85mm]{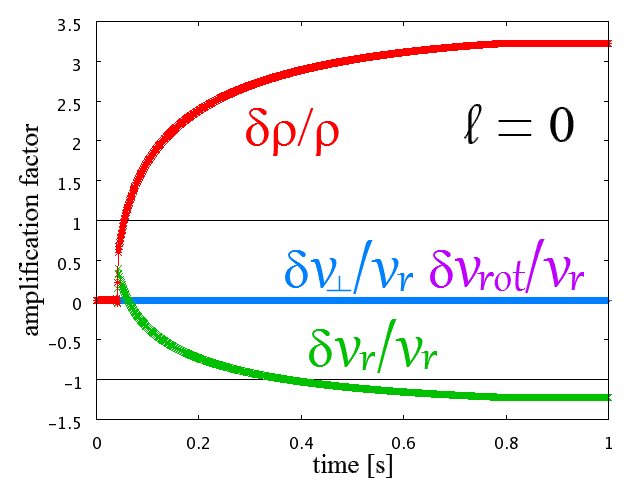}
\caption{The time evolutions of the $l = 0$ perturbations at the radius of $300$ $\mathrm{km}$. The step-function type perturbations are imposed at the outer boundary. The vertical axis is the amplification factors, i.e. the ratio to the values set at the outer boundary. The red, green, blue, and purple lines represent the perturbations of density, radial-velocity, and transverse components of velocity, respectively. Two horizontal black lines represent $\pm 1$.}
\label{fig.time evolution}
\end{figure}

We begin with spherical modes with $l=0$. In this case, the transverse components of velocity vanish everywhere at any time. We present the result in Fig.~\ref{fig.time evolution}, in which the vertical axis is the perturbations normalized by the values at the outer boundary whereas the horizontal axis represents the time from the instance when the perturbation is imposed at $r= R$. As seen in the figure, the perturbations reach $r_{sh}$ with a delay, which is estimated as
\begin{equation}
t_1 = \int _R^{r_{sh}}\frac{\diff r'}{\lambda (r')} = 41.5\; \mathrm{ ms},
\end{equation}
where $\lambda (r)$ denotes the velocity of the in-going acoustic wave: $v_r -c_s$.
Since the background flow is supersonic everywhere, the other two characteristic velocities, $v_r +c_s$ and $v_r$, are also negative.
It is also evident in the figure that the perturbations become steady after $t_3 = 783$ $\mathrm{ms}$, which corresponds to the time, at which a wave that has the slowest characteristic velocity, $v_r +c_s$, reaches $r_{sh}$. 
The perturbation increases monotonically in the case of density. It is doubled quickly in less than $200$ $\mathrm{ms}$ and nearly tripled finally in the steady state. The radial velocity grows much slowly and the amplification factor reaches only $\sim 1.25$ in the steady state.

\begin{figure}[htb]
\includegraphics[bb = 0 0 1254 1438, width = 85mm]{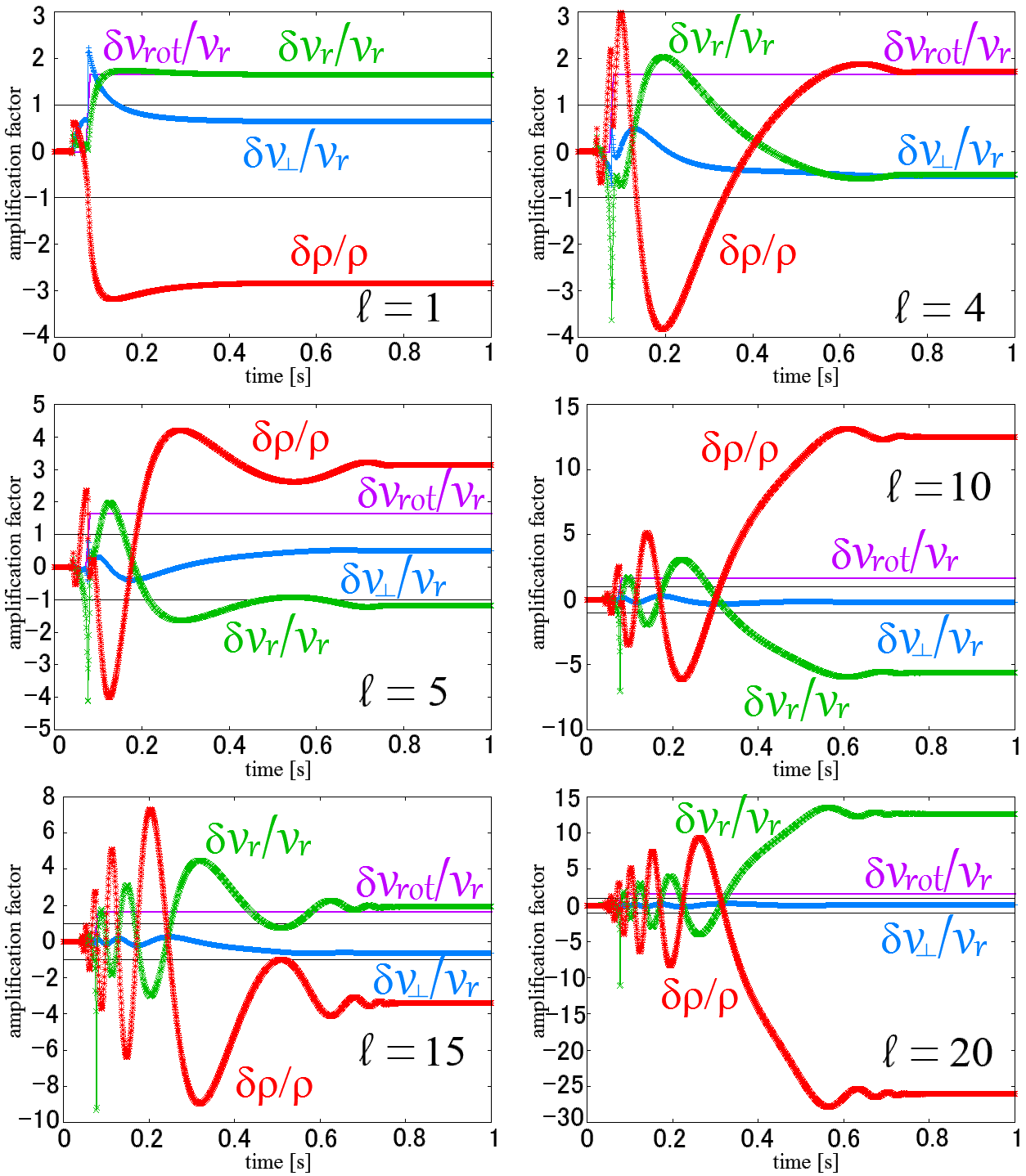}
\caption{The same as Fig.~\ref{fig.time evolution} but for different $l$'s.}
\label{fig.time evolution-others}
\end{figure}

Fig.~\ref{fig.time evolution-others} represents the results for $l = 1,4,5,10,15$, and $20$. For $l=20$, for example, the amplification factor reaches $\sim 30$ for the density perturbation and it goes up to $\sim 15$ for the radial-velocity perturbation in the steady state established after $t_3$. For these modes, the transverse components of velocity are also perturbed as $\delta v_\perp/v_r = \delta v_{rot}/v_r \propto \theta (t)$ at the outer boundary. As mentioned in Sec.~\ref{basic eq}, $\delta v_{rot}$ is not coupled with other perturbations and is obtained by Eq.~(\ref{analy}). As a matter of fact, for the step-function type perturbation assumed in this section, $\delta v_{rot}/v_r$ also becomes a step function in time with a discontinuity at $t = t_2$ given by
\begin{equation}
t_2 = \int _R^{r_{sh}}\frac{\diff r'}{v_r (r')}.
\end{equation}

\begin{figure}[htb]
\includegraphics[bb = 0 0 1217 1434, width = 85mm]{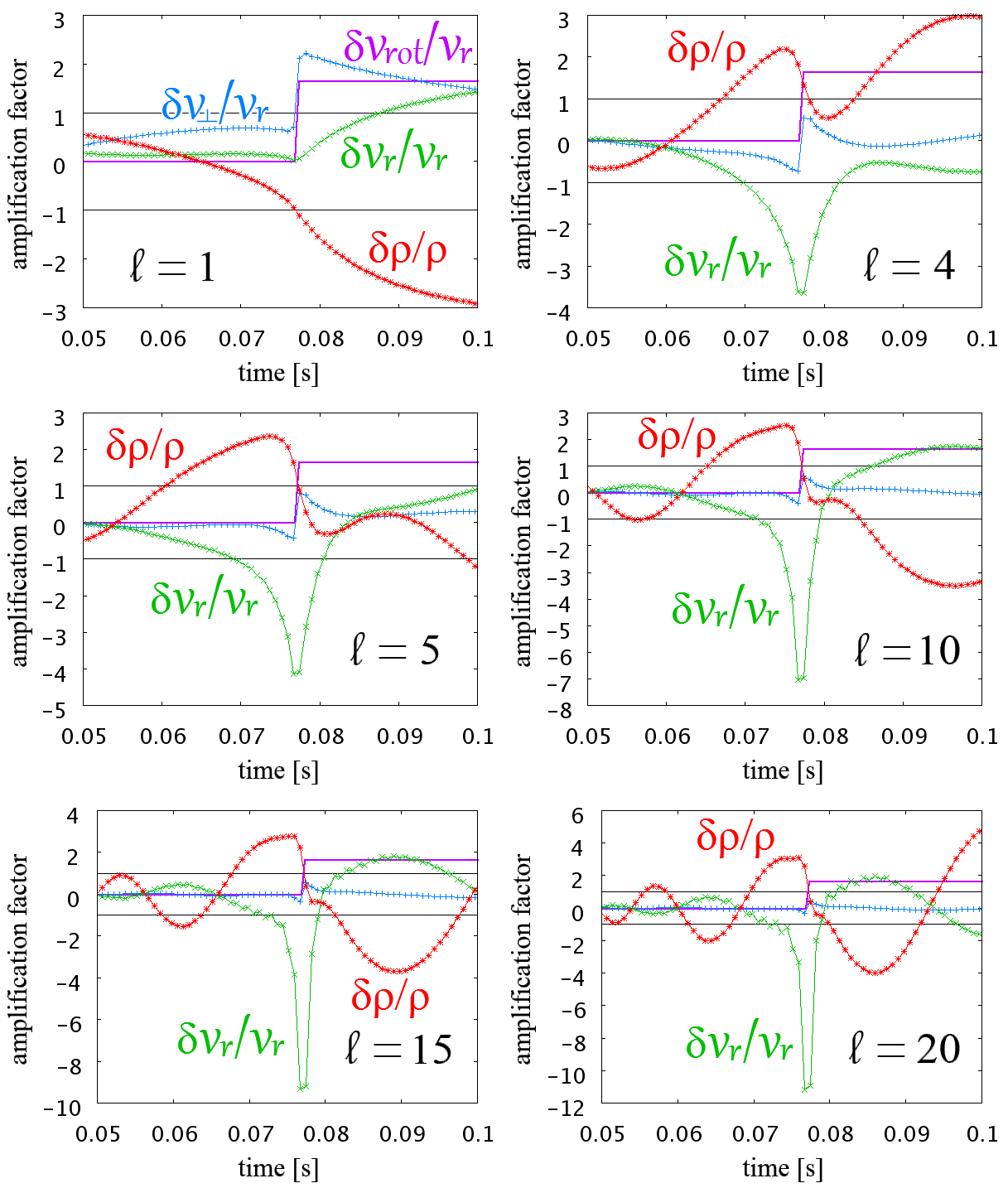}
\caption{The close-ups of Fig.~\ref{fig.time evolution-others} around $t=t_2$.}
\label{fig.close-ups}
\end{figure}

The time evolutions of other variables, on the other hand, change qualitatively as $l$ increases. Firstly, in contrast to the $l =0$ case, the density and radial-velocity perturbations have another transition at $t_2$ (see Fig.~\ref{fig.close-ups} for close-ups). This may seem strange, since the eigenfunction that corresponds to the eigenvalue, $v_r$, contains only $\delta v_\perp$ and hence $\delta \rho$ and $\delta v_r$ appear to have nothing special at $t_2$. This is not true, however, and these modes are actually mixed because the background flow is non-uniform spatially and, as a consequence, the eigenvectors vary radially.

Secondly, the perturbations oscillate in time between $t_1$ and $t_3$. Although these oscillations are not harmonic, their frequencies are roughly in the range of $40-100$ $\mathrm{s^{-1}}$ for $l = 1-20$ between $t_1$ and $t_2$ and they become higher as $l$ increases. The oscillations continue after $t_2$ but the frequencies get lower as the time passes: the interval between nodes are $14,19,25,35,51$, and $93$ $\mathrm{ms}$ for $l = 20$ while those for $l=10$ are $32,58$, and $128$ $\mathrm{ms}$. In general, the number of nodes and the intervals between them are larger and shorter, respectively, for greater $l$'s. 

Thirdly, as $l$ becomes larger, the amplitudes of the density and radial-velocity perturbations tend to get larger. It is analytically shown in Appendix~\ref{appB} that for large $l$'s the saturated amplitudes are proportional to $l$ for density and radial-velocity perturbations while those for the transverse velocities are independent of $l$. Note that these dependences are different from those claimed by LG00 or \citet{Kovalenko} for the asymptotic regime $(r\rightarrow 0)$. These authors assumed that the amplitudes obey a power law of $r$ in this asymptotic regime $(r\rightarrow 0)$ and deduced the dependence of $\propto l^2$.

\begin{figure}[htb]
\includegraphics[bb = 0 0 1296 1478, width = 85mm]{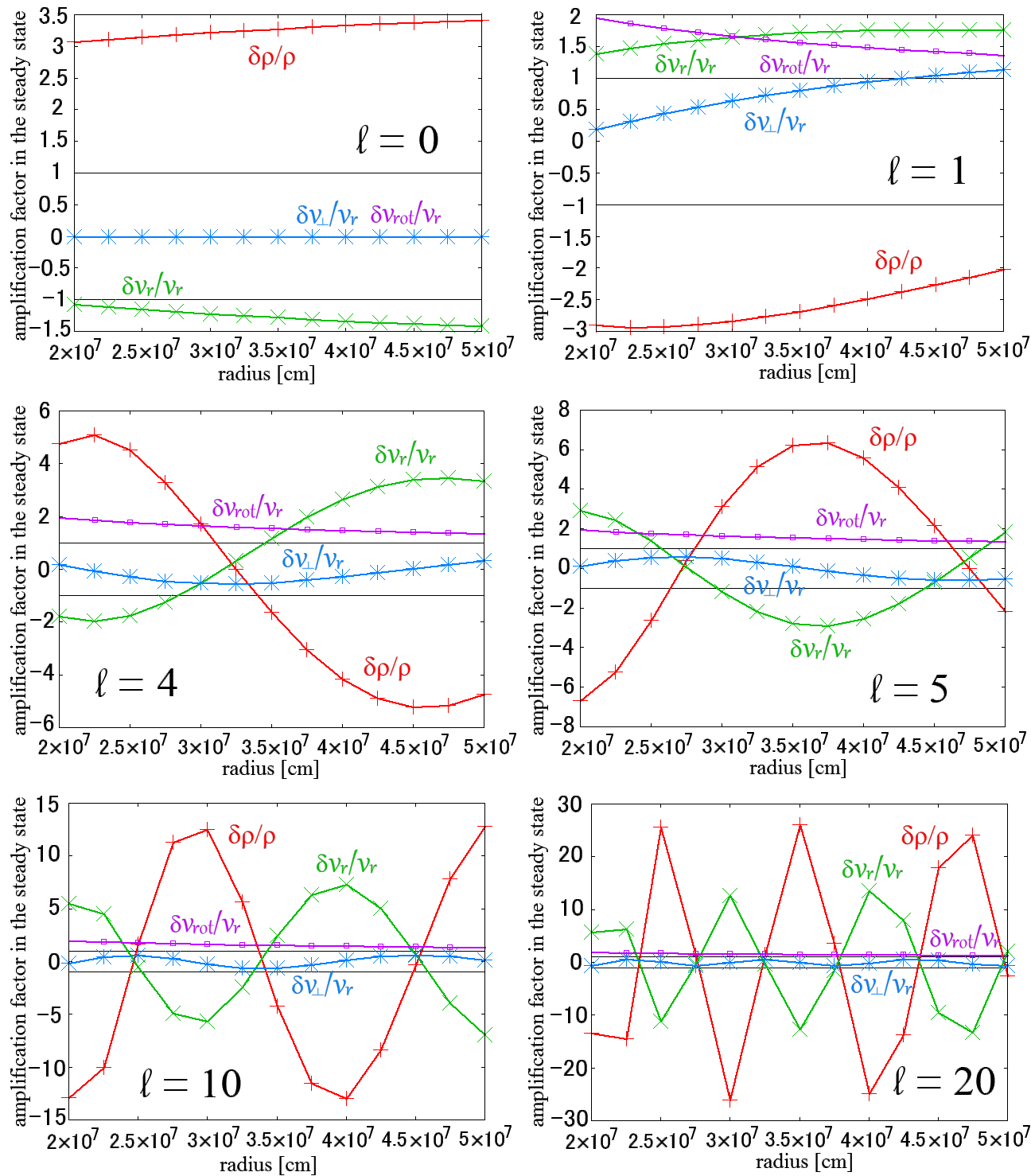}
\caption{The amplification factors in the steady state as a function of radius for various $l$'s. The red, green, blue, and purple lines represent the perturbations of density, radial-velocity, and transverse components of velocity, respectively. Two horizontal black lines represent $\pm 1$.}
\label{fig.sat_dist}
\end{figure}

So far we have fixed the inner boundary to $r = 300$ $\mathrm{km}$. We turn our attention to the dependence of the amplification factors on this radius. In Fig.~\ref{fig.sat_dist}, we plot the normalized perturbation amplitudes in the steady state at $t \geq t_3$ as a function of radius. As is evident, there are nodes in general, whose number becomes larger as $l$ increases, and the perturbation growth in space is never described by a simple power law of radius as assumed in the previous analyses by \citet{Kovalenko} and LG00. The radial variation may be better approximated by a sinusoidal wave. Note, however, that we are not dealing with the asymptotic regime $(r\rightarrow 0)$ unlike these authors. It is important that the amplitudes of the radial oscillations are largest for the density perturbation and those for the radial-velocity perturbation is second largest. More quantitative discussions on the radial oscillations are found in Appendix~\ref{appB}. 

\begin{figure}[htb]
\includegraphics[bb = 0 0 1292 1454, width = 85mm]{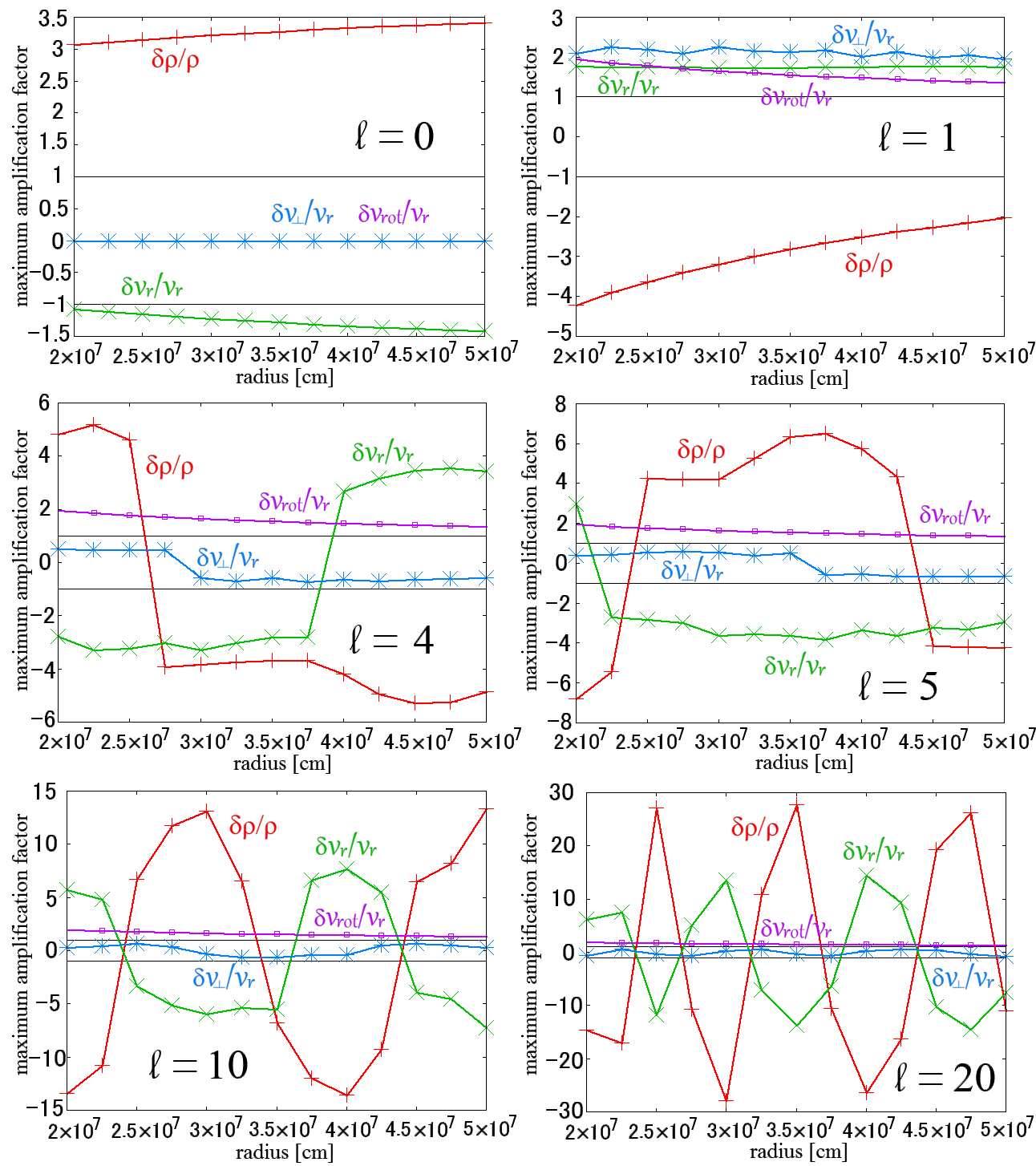}
\caption{The maximum amplification factors as a function of radius for different $l$'s. The red, green, blue, and purple lines represent the perturbations of density, radial-velocity, and transverse components of velocity, respectively. Two horizontal black lines represent $\pm 1$.}
\label{fig.max_dist}
\end{figure}

We also plot the maximum amplification factors, which are not necessarily attained in the steady state, at each radius in Fig.~\ref{fig.max_dist}. The lines for $\delta v_{rot}/v_r$ are identical in Figs.~\ref{fig.sat_dist} and \ref{fig.max_dist} for the reasons mentioned above. The plots for other variables are rather complicated and radial variations are sometimes rectangular rather than sinusoidal. For larger $l$'s, however, Figs.~\ref{fig.sat_dist} and \ref{fig.max_dist} become similar to each other, indicating that the maximum amplification factor is attained after the steady state is reached in these cases.

We note finally that the above features are not altered both qualitatively and quantitatively even if only $\delta v_\perp$ is non-vanishing at the outer boundary, which is the situation considered in \citet{Couch}. This is because the mode mixing explained earlier produces the density and radial-velocity perturbations. 

\subsection{sinusoidal perturbations} \label{result.sin}
\begin{figure}[htb]
\includegraphics[bb = 0 0 1255 1449, width = 85mm]{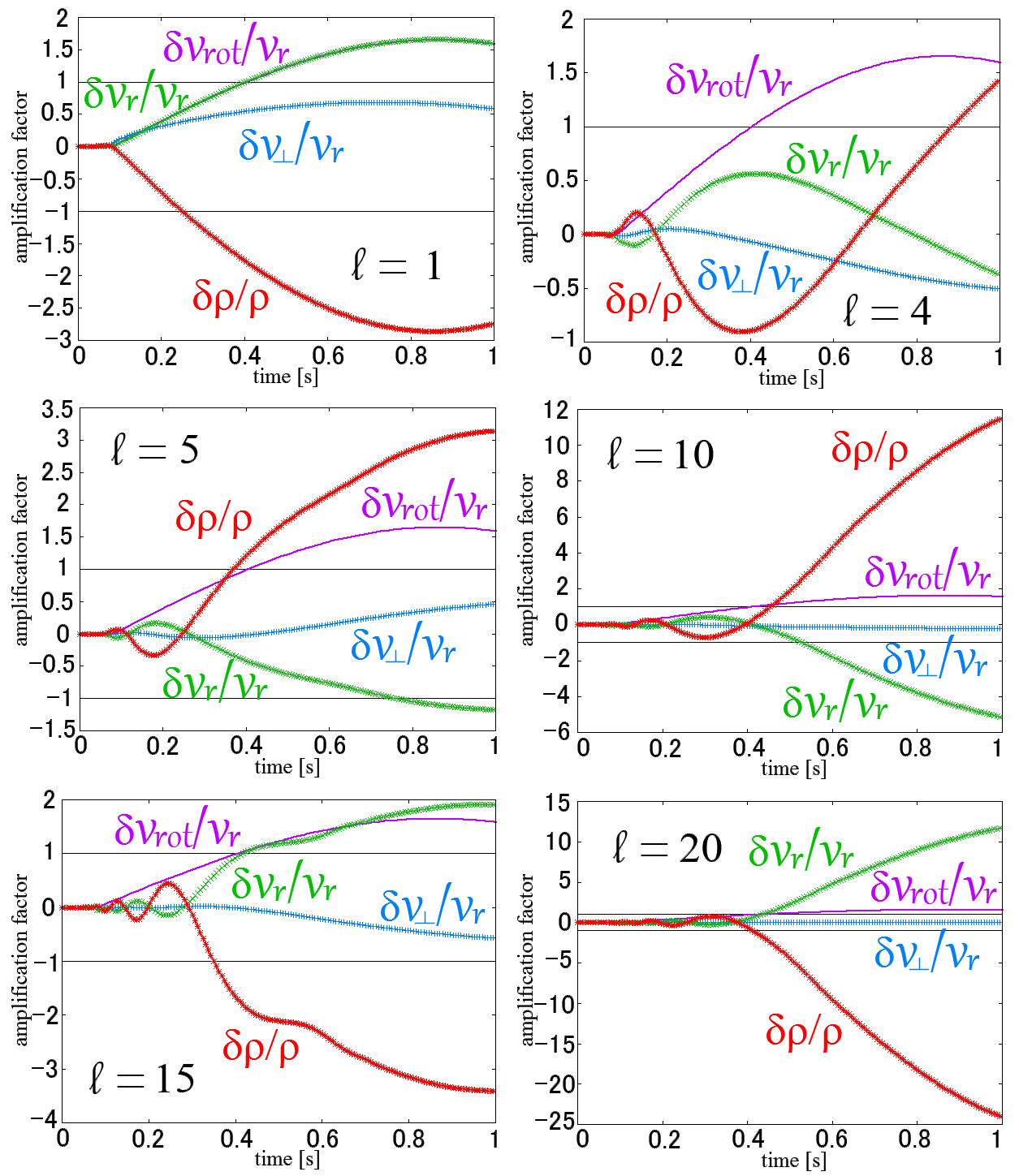}
\caption{The same as Fig.~\ref{fig.time evolution-others} but for the sinusoidal perturbations imposed at the outer boundary.}
\label{fig.sin}
\end{figure}

Although it facilitates the interpretation of results, the step-function type boundary condition adopted in the previous section is admittedly artificial. In this section we consider the models, in which a sinusoidal variation: $\delta \rho/ \rho = \delta v_r/ v_r = \delta v_\perp/v_r = \delta v_{rot}/v_r \propto \sin (\omega t)$, is assumed in the outer boundary condition. Here we set the angular frequency as $\omega  = 2$ $\mathrm{s^{-1} }$, which is approximately the inverse of the sound crossing time in the Si layer of the $15$ $\mathrm{M_\odot}$ progenitor.  The background model is not changed from the one in the previous section. We have repeated the same analysis for this model. We present here only the time evolutions of normalized perturbation amplitudes at $r_{sh} = 300$ $\mathrm{km}$ in Fig.~\ref{fig.sin}. Much like in the previous case, the perturbations oscillate rapidly in the early phase and grow later over the timescale of $\sim \omega ^{-1}$. In contrast to the step-function type perturbations, however, perturbations at a fixed point do not become steady but oscillate in a harmonic manner after $t_3$.

\begin{figure}[htb]
\includegraphics[bb = 0 0 1276 1446, width = 85mm]{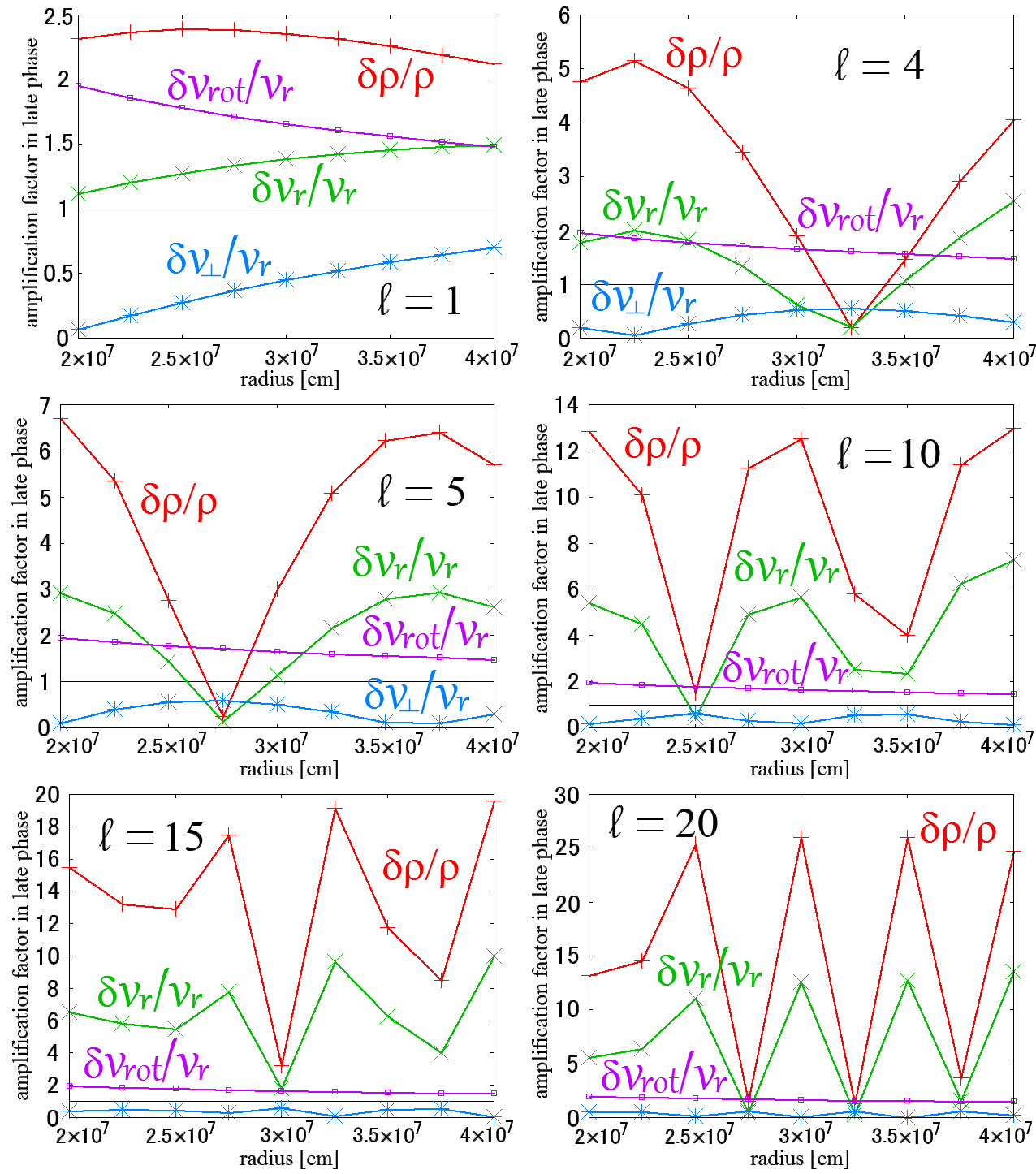}
\caption{The magnitudes of amplification factors in the late phase as a function of radius for various $l$'s. The red, green, blue, and purple lines represent the perturbations of density, radial-velocity, and transverse components of velocity, respectively. The horizontal black line represents unity.}
\label{fig.abs}
\end{figure}

We now turn our attention to the dependence of the amplification factors on the radius of the inner boundary. Fig.~\ref{fig.abs} shows the $r$-dependence of the magnitudes of amplification factors in the late phase, when the perturbations attain the harmonic oscillations at the inner boundaries. We again observe the same features as those in the previous section: the perturbation growth in space is never fitted by a simple power law of radius and the magnitudes of amplification factors are largest for density and those for radial velocity is second largest.

Although we have not investigated them, we believe that these features will be unchanged for other $\omega $.

\section{Discussions}\label{sec.discussion}
\subsection{comparisons with the previous results}
LG00 investigated the linear stability of supersonic accretion flows both analytically and numerically. In their analytic treatment, they considered only the asymptotic region, $r\rightarrow 0$, assuming a simple power law for the growth of perturbations. It is stressed that we are not dealing with this regime in this paper. In fact, since we have in mind the application of the results to the post-bounce phase of CCSNe, the inner boundary in this paper roughly corresponds to the stagnant shock wave in the supernova core and is rather distant $(\sim 100\ \mathrm{km})$ from the center. We should be aware of this difference in the following comparison. It is also noted that we neglect self-gravity whereas it was included in LG00.\footnote[4]{As already mentioned, the mass of accreting matter is less than 10 per cent of the central accretor in our canonical model. It is hence expected that the following results may be subject to change as much if self-gravity is taken into account.} As mentioned already, we do not find the power-law behavior in our model. We observe instead the oscillations in the spatial profile of the perturbation amplitudes. In the following we show that the envelopes of these oscillations obey power laws (see also Appendix~\ref{appB}). Even in that case, however, the powers we obtain are different from those reported in the previous papers as demonstrated below.

We first give the result of LG00 shortly. From Eqs.~(7), (22)-(25) in their paper, the perturbation growths are obtained in the asymptotic limit $(r\rightarrow 0)$ as
\begin{eqnarray}
\label{Lai1}
\frac{\delta \rho}{\rho } &\propto &2l(l+1)\frac{\delta v_\perp}{v_r} \propto 2l(l+1)r^{-1/2},\\
\label{Lai2}
\frac{\delta v_r}{v_r} &\propto &r^{-(3\gamma -4)/2} ,\\
\label{Lai3}
\frac{\delta v_\perp }{v_r} &\propto & r^{-1/2},\\
\frac{\delta v_{rot}}{v_r} &\propto & r^{-1/2},
\end{eqnarray}
under the assumption that the flow velocity in the background is $v_r\propto r^{-1/2}$.\footnote[5]{They further imposed the irrotational condition $\delta v_T := \delta v_r -\partial (r \delta v_\perp)/\partial r \equiv 0$, which is justified since the right hand side is proportional to $\sqrt{r}$ as $r\rightarrow 0$ anyway.}
Applying these relations to the entire region, we obtain the amplification factors as $8, 1.8, 2$, and $2$ for $\delta \rho/ \rho, \delta v_r/v_r, \delta v_\perp/v_r$, and $\delta v_{rot}/v_r$, respectively, if we adopt $R = 1.25\times10^8$ $\mathrm{cm}$, $r_{sh} = 3\times 10^7$ $\mathrm{cm}$, $l=1$ and $\gamma = 1.6$.

\begin{figure}[htb]
\includegraphics[bb = 0 0 640 480, width = 85mm]{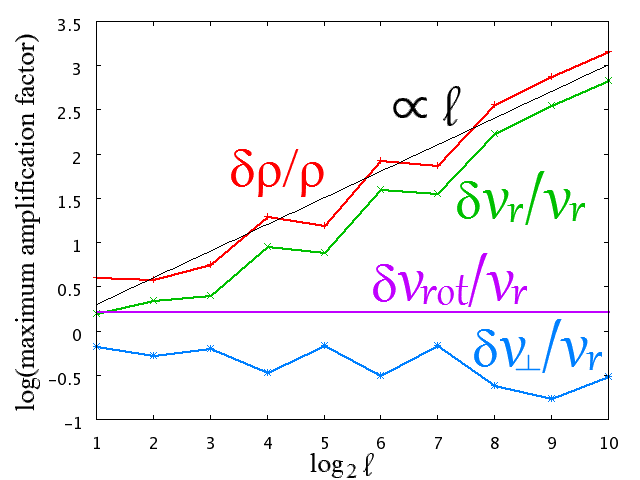}
\caption{The maximum amplification factors as a function of $l$. The step-function type perturbations are imposed at the outer boundary. The vertical axis is in the common logarithmic scale while the horizontal axis is in the base-2 logarithmic scale. The red, green, blue, and purple lines represent the perturbations of density, radial-velocity, and transverse components of velocity, respectively. The black line is a line of $\propto l$.}
\label{fig.LDependence}
\end{figure}

Now we use equations (\ref{analy}) and (\ref{y1})-(\ref{y2}) in Appendix~\ref{appB} to obtain the perturbation amplitudes for large $l$'s after the steady state is established. Note that the perturbation amplitudes in this steady state are not very different from the maximum values. Assuming that the velocity is proportional to $r^{-1/2}$ as LG00 did in the asymptotic limit of $r\rightarrow  0$, we obtain $\M  \propto r^{-(5 -3\gamma )/4}$ in the Bondi accretion flow. Inserting these relations to Eqs.~(\ref{analy}) and (\ref{y1})-(\ref{y2}), we have the following.
\begin{eqnarray}
\label{Ours1}
\frac{\delta \rho}{\rho } &\propto & lr^{-(5-3\gamma)/4} ,\\
\label{Ours2}
\frac{\delta v_r}{v_r} &\propto & lr^{(5-3\gamma )/4} ,\\
\label{Ours3}
\frac{\delta v_\perp }{v_r} &\sim &\mathrm{const.} ,\\
\frac{\delta v_{rot}}{v_r} &\propto & r^{-1/2}.
\end{eqnarray}
It should be mentioned that Eqs.~(\ref{y1})-(\ref{y2}) are obtained under the assumption that the Mach number is almost constant and hence are not applicable to the present case with $\M  \propto r^{-(5 -3\gamma )/4}$, rigorously speaking. However, the $r$-dependence of $\M$ is rather weak and the local application of Eqs.~(\ref{y1})-(\ref{y2}) may be justified. As a matter of fact, the $l$-dependence obtained this way reproduces the numerical results fairly well as shown in Fig.~\ref{fig.LDependence}.

It is evident from the comparison with Eqs.~(\ref{Lai1})-(\ref{Lai3}) that both the power and $l$-dependence are different. In fact, the power for $\delta \rho /\rho$, $-(5-3\gamma)/4$, is larger than $-1/2$ expected in LG00 as long as $\gamma > 1$, which is expected in CCSNe. Taking into account the difference of the $l$-dependence, in addition, it turns out that the amplification factor at the inner boundary assumed in the paper will be much smaller than expected by the previous study for large $l$'s.
As for $\delta v_r /v_r$, since the power in Eq.~(\ref{Ours2}) is positive for $\gamma < 5/3$ and larger than that in Eq.~(\ref{Lai2}) for $\gamma >1$, $\delta v_r /v_r$ will be decreased faster than supposed in LG00. The reason for the decrease in the perturbation amplitude is just more rapid increases in the background velocity. Note that this is true only near the inner boundary and $\delta v_r/v_r $ increases at the beginning. Combined with the $l$-dependence, the amplification factor can become much larger than unity as shown in Fig.~\ref{fig.max_dist}. Finally, $\delta v_\perp/v_r$ is not amplified according to Eq.~(\ref{Ours3}), which is in sharp contrast to the prediction in LG00 that it is inversely proportional to the square root of $r$. Fig.~\ref{fig.max_dist} demonstrates that it is indeed smaller than unity for large $l$'s.

It is finally mentioned that \citet{Kovalenko} also investigated the stability of Bondi accretion flows in the asymptotic regime and obtained $\delta v_r \propto l^2$, which is different both from our result and from the LG00. On the other hand, they found the same $l$-dependence for the density perturbation as LG00 did, which is different from ours as stated above.

\subsection{possible impact on the shock dynamics in CCSNe}
The above analysis suggests that the perturbations generated in the outer envelope of a massive progenitor may be amplified by an order by the time when they reach the stalled shock wave. This implies that the initial perturbation amplitudes of a few percent may be sufficient to affect shock revival. Such a number may indeed obtain in violent convections that Arnett and his company advocated \citep{Bazan, Asida, Meakin06, Meakin07, Arnett}. As demonstrated above, since the density and radial-velocity perturbations are amplified in proportion to $l$, the spectrum of the initial perturbation is important to identify the dominant $l$. It was recently studied by \citet{Chat} in their multi-dimensional simulations of oxygen shell burnings of a $15$ $\mathrm{M_\odot}$ progenitor. They reported that the power spectrum peaks at $l = 8 \; (5)$ and then decays exponentially for the two- (three-) dimensional case. It is then expected from our study that the dominant modes that will affect most the stalled shock wave will be also those with $l \sim 8 (5)$.

It is also intriguing to point out that the perturbations will oscillate in time at the shock wave with frequencies of $10-10^2$ $\mathrm{s^{-1}}$, which are rather close to the typical frequencies of SASI \citep[e.g.][]{Iwakami}. This similarity might play some role in reviving the stagnated shock wave in \citet{Couch}. Further investigations are certainly warranted (Takahashi \& Yamada 2014, in preparation).

\section{Summary }\label{sec.summary}
We have studied the linear growth of the perturbations that are generated at a large radius and propagate inward in a spherically symmetric supersonic accretion flow, having in mind the application to the investigation of shock revival in CCSNe. In contrast to the previous studies, we have solved the linearized equations as an initial and boundary value problem, employing Laplace transform, which enables us to obtain the amplification factor at a specified point easily.

The background flow is chosen to be a supersonic portion of a transonic Bondi accretion flow whose parameters are set to mimic the collapse of a supposedly typical supernova progenitor: the $15$ $\mathrm{M_\odot}$ star of \citet{Woosley}.
We have considered two seed perturbations: a step-function type perturbation and a sinusoidal one. The former is more experimental and meant to elucidate the systematics.  

We have found that the density and radial-velocity perturbations grow as they propagate inward. In fact, the amplification factors can be more than $10$ when applied to the perturbations generated in the Si/O layer of the massive progenitor and accreted on the stalled shock wave in the CCSN core. They are actually oscillatory in both time and space, which is in sharp contrast to the previous studies that predicted power-law behavior in the asymptotic limit $(r\rightarrow 0)$. We have shown analytically that the envelopes of the oscillatory amplification factors may obey power laws for large $l$'s. We have observed, however, that the powers are still different from the previous results. We have also demonstrated both analytically and numerically for large $l$'s that the amplification factors for the density and radial-velocity perturbations are proportional to $l$, which is again at odds with the previous expectations. These discrepancies are most likely due to the difference in the regimes of supersonic flows, however. The previous works investigated the innermost part of the Bondi flow $(r\rightarrow 0)$, in which the Mach number becomes high and varies rapidly. In this paper, on the other hand, we have studied the outer part of the same flow, at which the Mach number is close to unity and changes rather slowly. The latter regime is more appropriate for the post-bounce phase of CCSNe.

We have found that the typical oscillation frequencies of perturbations at the stalled shock wave in the CCSN core are not much different from the canonical frequencies of SASI. This may have an important implication for shock revival and will be our next target. The results of linear analysis will be reported in our forthcoming paper (Takahashi \& Yamada 2014, in preparation).

\acknowledgments
KT thanks Yu Yamamoto for discussions on the importance of the progenitor dependence of CCSNe. 
He is also thankful to Ko Nakamura and Wakana Iwakami for discussions on SASI dynamics in their numerical investigations. This work is partially supported by Research Fellowship for Young Scientists from the Japan Society for the Promotion of Science as well as by the Grants-in-Aid for the Scientific Research (A) (NoS. 24244036, 24740165), the Grants-in-Aid for the Scientific Research on Innovative Areas, “New Development in Astrophysics through multi messenger observations of gravitational wave sources” (No. 24103006).

\appendix
\section{The integration path in the inverse Laplace transform}\label{appA}
Here we consider the integral path in the inverse Laplace transform, that is, how to choose $x$ in Eq.~(\ref{inverse_x}) (see also Sec.~\ref{Laplace method}). The value of $x$ should be chosen so that the entire path would lie in the region, in which $f^*(s)$ exists, i.e. the right-half ($\mathrm{Re}[ s] > s_0$ for a certain real number $s_0$) of the complex $s$-plane. Hence we need to find $s_0$ first. The Laplace transformed basic equations (\ref{Lrho})-(\ref{Lperp}) can be expressed as follows:
\begin{equation} \label{matrix_form}
\frac{\diff {\bf y}^*}{\diff  r} = [sA(r) +B(r)]{\bf y}^*,
\end{equation}
where ${\bf y} ^*={\bf y} ^*(r,s)$ is a vector whose components are the Laplace transformed variables: ${\bf y}^* := (\delta \rho^*/\rho ,\ \delta v_r^*/v_r,\ \delta v_\perp ^*/v_r)$. The matrices $A(r) $ and $B(r)$ are defined as 
\begin{eqnarray}\label{A}
A(r) &=& \frac{1}{c_s(1-\M ^2)} \left(
\begin{array}{ccc}
\M & -\M & 0 \\
\displaystyle -\frac{1}{\M} & \M & 0 \\
0 & 0 & \displaystyle \frac{\M ^2 -1}{\M }
\end{array} \right), \\ \label{B}
B(r) &=& \frac{1}{1-\M ^2} \left(
\begin{array}{ccc}
\displaystyle \M^2  \frac{\diff}{\diff r}\ln \dot{M}  -\frac{\gamma -1}{\rho}\frac{\diff \rho}{\diff r} & 
\displaystyle \M^2 \left( \frac{\diff}{\diff r}\ln \dot{M} -\frac{2}{v_r}\frac{\diff v_r}{\diff r} \right) & 
\displaystyle -\frac{\M ^2 l(l+1)}{r} \\
\displaystyle -\frac{\diff}{\diff r}\ln \dot{M} +\frac{\gamma -1}{\rho }\frac{\diff \rho}{\diff r} &
\displaystyle -\frac{\diff}{\diff r}\ln \dot{M} +\frac{2\M^2}{v_r}\frac{\diff v_r}{\diff r} & 
\displaystyle \frac{l(l+1)}{r} \\
\displaystyle \frac{\M ^2 -1}{r\M ^2} & 0 & 
\displaystyle (\M ^2 -1)\left(\frac{1}{v_r}\frac{\diff v_r}{\diff r} + \frac{1}{r} \right)
\end{array} \right).
\end{eqnarray}
Note that $\diff \ln \dot{M}/\diff r = 0$ in steady accretions. 
Equation (\ref{matrix_form}) is solved formally:
\begin{equation}
{\bf y}^*(r,s) = \mathcal{P} \left[ \exp \left[ \int _R ^r \diff r' \left( sA(r') +B(r') \right) \right] \right] {\bf y}^* (R, s),
\end{equation}
where $\mathcal{P}$ is an operator that takes a path-ordered product \citep{Peskin}. The norm of the solution is then estimated as follows.
\begin{eqnarray}
|| {\bf y}^*(r,s) || &\le& \left| \left| \mathcal{P} \left[ \exp \left[ \int _R ^r \diff r' \left( sA(r') +B(r') \right) \right] \right]\right| \right| || {\bf y}^*(R,s)|| ,\\
&\le& \sum _{k=0}^\infty \frac{1}{k!} \int_R^r \cdots \int_R^r |\diff r_1'| \cdots |\diff r_k'| \nonumber \\
& &\; \; \times ||\mathcal{P}[(sA(r_1') +B(r_1'))\cdots (sA(r_k') +B(r_k'))]||\  || {\bf y}^*(R,s)||.
\end{eqnarray}
Since the background flow is non-singular and we are concerned with a finite interval of $r$, the norm of the matrix $sA(r)+B(r)$ is bounded as
\begin{eqnarray}
||sA(r) +B(r)|| &\le & ||sA(r)|| + ||B(r)||, \\
&\le & |s|C_A + C_B,
\end{eqnarray}
where $C_A$ and $C_B$ are some positive constants. And $|R-r|$ is also bounded by a constant, say, $L$. We hence obtain the following:
\begin{eqnarray}
|| {\bf y}^*(r,s) || &\le& \sum _{k=0}^\infty \frac{1}{k!} \int_R^r \cdots \int_R^r |\diff r_1'| \cdots |\diff r_k'|\ (|s|C_A +C_B)^k  || {\bf y}^*(R,s)||, \\
&\le& \sum _{k=0}^\infty \frac{L^k(|s|C_A+C_B)^k }{k!}|| {\bf y}^*(R,s)||, \\
&=& \exp \left[ L (|s|C_A+C_B) \right] || {\bf y}^*(R,s)||.
\end{eqnarray}
Since $\exp \left[ L (|s|C_A+C_B) \right]$ is finite for any complex $s$, the singularity of the Laplace transformed solution, if any, should originate from the boundary function, ${\bf y}^*(R,s)$. From this fact, we know \textit{a priori} where the singularities exist and hence how to choose $x$.
For example, if a sinusoidal perturbation, $\sin(\omega t)$, is imposed at the outer boundary, the Laplace transformed solution has two poles at $s = \pm i\omega$, since $\Laplace [\sin(\omega t)] = \omega /(s^2 +\omega ^2)$. In this case, we can choose any positive real $x$, i.e. $x>0$.

We could not choose so large an $x$, however. Recalling the formula (\ref{inverse2}) and the fact that the inverse Laplace transform is independent of $x$, we know that
\begin{equation}
\int _{-\infty} ^\infty f^*(x,y)e^{iyt}\diff y \propto e^{-tx},
\end{equation}
since the prefactor $\exp[tx]$ in Eq.~(\ref{inverse2}) should be canceled out. In addition to the requirement that these factors should not overflow or underflow, we need to ensure the cancellation numerically, which would be difficult for large values of $x$.

On the other hand, we had better not choose so small an $x$, either. In fact, as $x$ becomes smaller, the integral path moves closer to the singularities and the vicinities of the singularities contribute more to the integral. Then, we need to deploy a finer grid for the integral, $\Delta y$, which would be numerically costly.

\section{Perturbation growths in steady states} \label{appB}
Here we derive time-independent solutions of Eqs.~(\ref{Lrho})-(\ref{Lperp}) in some limits, which would be useful in understanding the systematics of perturbation growths. They are obtained formally by taking the $t\rightarrow \infty$ limit, which in turn corresponds to the limit of $s\rightarrow 0$ in the Laplace transformed quantities. We assume that $l$ is sufficiently large and consider only its leading terms. Unlike the previous works we do not take the limit of $r\rightarrow 0$ here. Instead the inner boundary is rather distant from the center. As a result, the Mach number is not very high and changes slowly. We further assume that ${\bf y }(r,t)= (\delta \rho/\rho ,\ \delta v_r/v_r,\ \delta v_\perp /v_r)$ grows with $l$ at most as $\propto l $, which is actually observed in the numerical results and will be also justified a posteriori. Then the equations that govern steady states ${\bf y} _\infty = {\bf y}(r, t=\infty)$ are
\begin{eqnarray}\label{steady}
\frac{\diff {\bf y}_\infty}{\diff  r} &=& B_\infty(r){\bf y}_\infty, 
\end{eqnarray}
in which the matrix is given as
\begin{eqnarray}
B_\infty (r) &=& \left(
\begin{array}{ccc}
0 & 0 & \displaystyle -\frac{\M ^2 l(l+1)}{(1-\M ^2)r} \\
0 & 0 & \displaystyle \frac{l(l+1)}{(1-\M ^2)r} \\
\displaystyle -\frac{1}{r\M ^2} & 0 & \displaystyle -\frac{1}{v_r}\frac{\diff v_r}{\diff r} -\frac{1}{r}
\end{array} \right).
\end{eqnarray}
From the above equations, we can write down the equation for $y_{\infty ,1} (= \delta \rho /\rho| _{t=\infty})$, the first component of ${\bf y}_\infty$, as
\begin{eqnarray}
\frac{\diff ^2 y_1}{\diff r^2} +\frac{\diff }{\diff r}\ln \left[ \frac{r^2v_r (\M^2 -1)}{\M^2}\right]\frac{\diff y_1}{\diff r} +\frac{l(l+1)}{r^2(\M ^2-1)}y_1 = 0.
\end{eqnarray}
In this equation and hereafter we omit the subscript $\infty$ for notational simplicity.
Suppose that $v_r$ is described by a power law of $r$ and $\M$ is constant $(=\M _0)$. Then the first factor of the second term is proportion to $1/r$, and the equation is rewritten with some constant, $\alpha$, as follows:
\begin{eqnarray}\label{eq.y1}
\frac{\diff ^2 y_1}{\diff r^2} +\frac{\alpha}{r}\frac{\diff y_1}{\diff r} +\frac{l(l+1)}{r^2(\M _0 ^2-1)}y_1 = 0.
\end{eqnarray}
Since $r=0$ is a regular singularity, Eq.~(\ref{eq.y1}) can be solved by the Frobenius method as follows.
\begin{eqnarray}
y_1(r) &=& y_1(r_0)\cos \left[ \frac{l \ln \left( r/r_0 \right) }{\sqrt{\M _0^2 -1}} \right] +\frac{r_0\sqrt{\M _0^2 -1}}{l}\left. \frac{\diff y_1}{\diff r}\right| _{r=r_0}\sin \left[ \frac{l \ln \left( r/r_0 \right) }{\sqrt{\M _0^2 -1}} \right], \\
&=& y_1(r_0)\cos \left[ \frac{l \ln \left( r/r_0 \right) }{\sqrt{\M _0^2 -1}} \right] + \frac{\M _0 ^2(l+1)}{\sqrt{\M _0^2 -1}}y_3(r_0) \sin \left[ \frac{l \ln \left( r/r_0 \right) }{\sqrt{\M _0^2 -1}} \right],\\ \label{y1}
&\sim & \frac{\M _0^2l }{\sqrt{\M _0^2 -1}}y_3(r_0) \sin \left[ \frac{l \ln \left( r/r_0 \right) }{\sqrt{\M _0^2 -1}} \right],
\end{eqnarray}
where $r_0$ is some reference radius and we used the approximation $l \rightarrow \infty$. From (\ref{y1}), we find that the amplitude of density perturbation is proportional to $l$ and its radial profile is approximated by a sinusoidal function of $\ln r$. Note that the argument is also proportional to $l$ and inversely proportional to $\M _0$ if $\M _0 \gg 1$.

Employing the same assumptions and approximations, we obtain
\begin{eqnarray} \label{y3}
y_3(r) \sim y_3(r_0)\cos \left[ \frac{l\ln\left(r/r_0 \right)}{\sqrt{\M _0 ^2 -1}} \right].
\end{eqnarray}
In contrast to the density perturbation, the perturbation of transverse velocity is independent of $l$. Its radial distribution is again a sinusoidal function of $\ln r$ although the phase is shifted by $\pi /2$ radian with respect to that of the density perturbation, which is indeed observed in Fig.~\ref{fig.sat_dist}.

For the radial-velocity perturbation, the following relation is obtained by integrating Eq.~(\ref{steady}),
\begin{eqnarray} \label{y2}
y_2(r) \sim -\frac{l y_3(r_0)}{\sqrt{\M _0^2 -1}}\sin \left[ \frac{l\ln \left( r/r_0\right)}{\sqrt{\M _0^2 -1}} \right] \sim -\frac{1}{\M _0^2} y_1(r).
\end{eqnarray}
Hence the radial-velocity perturbation is simply proportional to the density perturbation but the amplitude is smaller by a square of the Mach number and the phase is also shifted by $\pi $ radian. These features are again discernible in Fig.~\ref{fig.sat_dist}.

\end{document}